\documentclass[namedreferences]{solarphysics}

\usepackage[hyperref,optionalrh,showbiblabels]{spr-sola-addons} % For Solar Physics 
\usepackage{graphicx}        % For eps figures, newer & more powerfull
\usepackage{amssymb}        % useful mathematical symbols
\usepackage{bm}
\usepackage{color}           % For color text: \color command
\usepackage{breakurl}        % For breaking URLs easily trough lines
\usepackage{hyperref}
\usepackage{gensymb}
\usepackage{multirow}
\usepackage{hyperref}
\usepackage[dvipsnames]{xcolor}

\newcommand{\Fig}[1]{Figure~\ref{#1}}

\newcommand{\aap}{    {\it Astron. Astrophys.}}

\newcommand{\aapr}{   {\it Astron. Astrophys. Rev.}}

\newcommand{\apj}{    {\it Astrophys. J.}}

\newcommand{\mnras}{  {\it Mon. Not. Roy. Astron. Soc.}}

\newcommand{\solphys}{{\it Solar Phys.}}

\newcommand{\al}{     {\it Astron. Lett.}}
\newcommand{\aspcs}{  {\it Astr. Soc. Pacific Conf. Ser.}}
\newcommand{\lrsp}{  {\it Living Rev. Sol. Phys.}}

\chardef\us=`\_

\begin{document}
\begin{article}
\begin{opening}

\title{Measurements of Solar Differential Rotation Using the Century Long Kodaikanal Sunspot Data}

%===========================================================================================================
%authors list
\author[addressref={aff1,aff2,aff6},corref,email={bibhuti.kj@iiap.res.in, maitraibibhu@gmail.com}]{\inits{B. K.}\fnm{Bibhuti Kumar}~\lnm{Jha}\orcid{0000-0003-3191-4625}}%\sep
\author[addressref={aff1,aff2}]{\inits{A.}\fnm{Aditya}~\lnm{Priyadarshi}\orcid{0000-0003-2476-1536}}%\sep
\author[addressref={aff3}]{\inits{S.}\fnm{Sudip}~\lnm{Mandal}\orcid{0000-0002-7762-5629}}%\sep
\author[addressref={aff4}]{\inits{S.}\fnm{Subhamoy}~\lnm{Chaterjee}\orcid{0000-0002-5014-7022}}%\sep
\author[addressref={aff1,aff2,aff5},corref,email={dipu@iiap.res.in}]{\inits{D.}\fnm{Dipankar}~\lnm{Banerjee}\orcid{0000-0003-4653-6823}}%\sep

%=========================================================================================================
%address
\address[id=aff1]{Indian Institute of Astrophysics, Koramangala, Bangalore 560034, India}
\address[id=aff2]{Aryabhatta Research Institute of Observational Sciences, Nainital 263000,Uttarakhand, India}
\address[id=aff6]{Pondicherry University, Chinna Kalapet, Kalapet, Puducherry 605014, India}
\address[id=aff3]{Max Planck Institute for Solar System Research, Justus-von-Liebig-Weg 3, D-37077 G\"ottingen,Germany}
\address[id=aff4]{Southwest Research Institute, 1050 Walnut Street \#300, Boulder, CO 80302, USA}
\address[id=aff5]{Center of Excellence in Space Sciences India, IISER Kolkata, Mohanpur 741246, West Bengal, India}

%=======================================================================================================
\runningauthor{B. K. Jha et. al.}
\runningtitle{Differential Rotation of the Sun Using KoSO Data}
\begin{abstract}
 The rotational profile of the Sun is considered to be one of the key inputs in a solar dynamo model. Hence, precise and long-term measurements of this quantity is important for our understanding of solar magnetism and its variability. In this study, we use the newly digitized, white light sunspot data (1923\,--\,2011) from Kodaikanal Solar Observatory (KoSO) to derive the solar rotation profile. An automated correlation based sunspot tracking algorithm is implemented to measure the rotation parameters, $A$, the equatorial rotation rate and $B$, the latitudinal gradient. Our measurements of $A=14.381\pm0.004$ and $B=-2.72\pm0.04$ compare well with previous studies. In our analysis, we find that the bigger sunspots (with area $>$400~$\mu$Hem) rotate slower than the smaller ones. At the same time, we do not find any variation in the rotation rates between activity extremes, i.e solar maxima and minima. Lastly, we employ our tracking algorithm on the Michelson Doppler Imager (MDI) data and compare the MDI results with our KoSO values.

\end{abstract}
\end{opening}
%======================Introduction==============================
\section{Introduction}
\label{s-intro}
A sunspot, a dark photospheric feature, is widely considered to be a suitable proxy of solar surface magnetism and its long-term variability. As we now understand, a dynamo which operates beneath the photospheric layer, is responsible for generating the magnetic field which we observe on the surface \citep{Pa55b}. Within this framework, a sufficiently strong and buoyant flux tube rises through the convection zone and forms a bipolar magnetic patch which often manifests itself as a spot pair on the visible solar surface \citep{Pa55b, Solanki2003}. Thus, analysis of spot properties on the surface, presents a unique opportunity to sneak peek the sub-surface physical processes which are otherwise hidden from us. 

One of the key solar parameters that can be measured using sunspots, is the differential rotation profile of the Sun. In fact, it is this differential rotation in the solar dynamo theory which stretches the poloidal field and converts it into the spot-generating toroidal field \citep{PC2010}. Therefore, precise and long-term measurements of solar rotation rates are required to further help and improve the current dynamo models \citep{badalyan_2017}. Modern day space-borne high-resolution data from the Solar and Heliospheric Observatory \citep[SOHO:][]{Domingo1995} and the Solar Dynamic Observatory \citep[SDO:][]{Pesnell2012} offer significant improvements in measuring the rotation profile. However, such data are only limited to the last two solar cycles (1996 onward). On the other hand, regular sunspots measurements from a number of ground based observatories are available for a significantly longer time (more than a century) and can be used to determine the differential rotation profile of the past.

Being one of the oldest recorded solar parameters, sunspots have already been utilised many times in the past to derive the rotation profile. Earliest of such measurements were reported by Christoph Scheiner in 1630 and almost after 200 years, by \citet{carrington1863}. These authors followed spots as they moved across the solar disc and reported a differential rotation profile in which the equator rotates faster than the poles. Later, as more sunspot data from different observatories became available, several follow-up studies on this subject were conducted. These include analyses using sunspot data from the Royal Greenwich Observatory \citep[][etc]{newton1951, ward1966, balthasar_1986, java2005}, Kenzelh\"ohe Observatory \citep{lustig_1983}, Mt. Wilson Observatory \citep[][etc.]{HG1984, GH1984, GH1985}, Meudon Observatory \citep{Ribes1993}, Catania Observatory \citep{ternullo_1981}, Kodaikanal Solar Observatory \citep[KoSO:][]{howard1999a, gupta1999}, etc. Overall, the observed rotation profile ($\Omega$) can be expressed by the following mathematical formula \citep{newton1951, HG1984}:

\begin{equation}
    \Omega~=~A~+~B\sin^2 {\theta}~+~C\sin^4{\theta}.
    \label{eq1}
\end{equation}

Since sunspots are mostly restricted within $\pm$40$\degree$ latitudes, one can drop the $C$ term \citep{newton1951, HG1984, wohl_2010} and rewrite the equation as:
 \begin{equation}
    \Omega~=~A~+~B\sin^2 {\theta},
    \label{eq2}
\end{equation}

where $A$ and $B$ represent the equatorial rotation rate and the latitudinal rotation gradient, respectively. Here, $\theta$ is the heliographic latitude of the spot.

Despite the considerable amount of independent studies on this subject, there still remain certain systematic differences in the rotation rates derived from different sunspot catalogues (see Table~\ref{table1}). This is partly due to the use of different spot tracing methods \citep{newton1951, HG1984,poljan2017}, different trace subjects within the spot (such as only the umbra vs. the whole spot; \citealp{howard1999a, gupta1999}) etc. In fact, in most of the older publications, authors had used manual inspections to identify the spots \citep{newton1951, gupta1999}. This, in turn, has introduced human errors and biases.

In this context, we present here the long-term measurements (1923\,--\,2011) of solar rotation rates using the newly digitised, high resolution KoSO sunspot observations. A fully automated sunspot tracing algorithm has been implemented to eliminate any detection bias and robustness of the method is also demonstrated by applying it on modern data from the Michelson Doppler Imager \citep[MDI:][]{scherrer1995} onboard SOHO.

%=========================Method================================
\section{ Data and Method}
\label{s-method}

Kodaikanal Solar Observatory (KoSO) is recording regular white-light sunspot data since 1904. This data, originally stored in photographic plates and films, have now been digitised into high resolution images of 4k$\times$4k size (for more information about this digitisation, see \citealp{ravindra2013}). Barring the initial 15 years (1904\,--\,1920), calibration as well as sunspot detection on the rest of the data (1921\,--\,2011) have also been completed recently \citep{ravindra2013, mandal2017}. KoSO sunspot area catalogue\footnote{Available for download at \url{https://kso.iiap.res.in/new/white_light}.} provides area values (i.e umbra+penumbra) of individual spots\footnote{Individual spots are not yet classified into groups in the current KoSO catalogue.} along with their heliographic positions (longitude and latitude).

For the purpose of this work, we utilise the binary, full-disc images of detected sunspots as generated by \citet{mandal2017} (see Section~4 of \citet{mandal2017} for more detalis). In this work, we implement a fully automated algorithm to track the spots in successive sunspot images. The whole procedure can be summarised as follows. 
%============================
\begin{figure}[!htbp]
\centerline{\includegraphics[width=\textwidth,clip=]{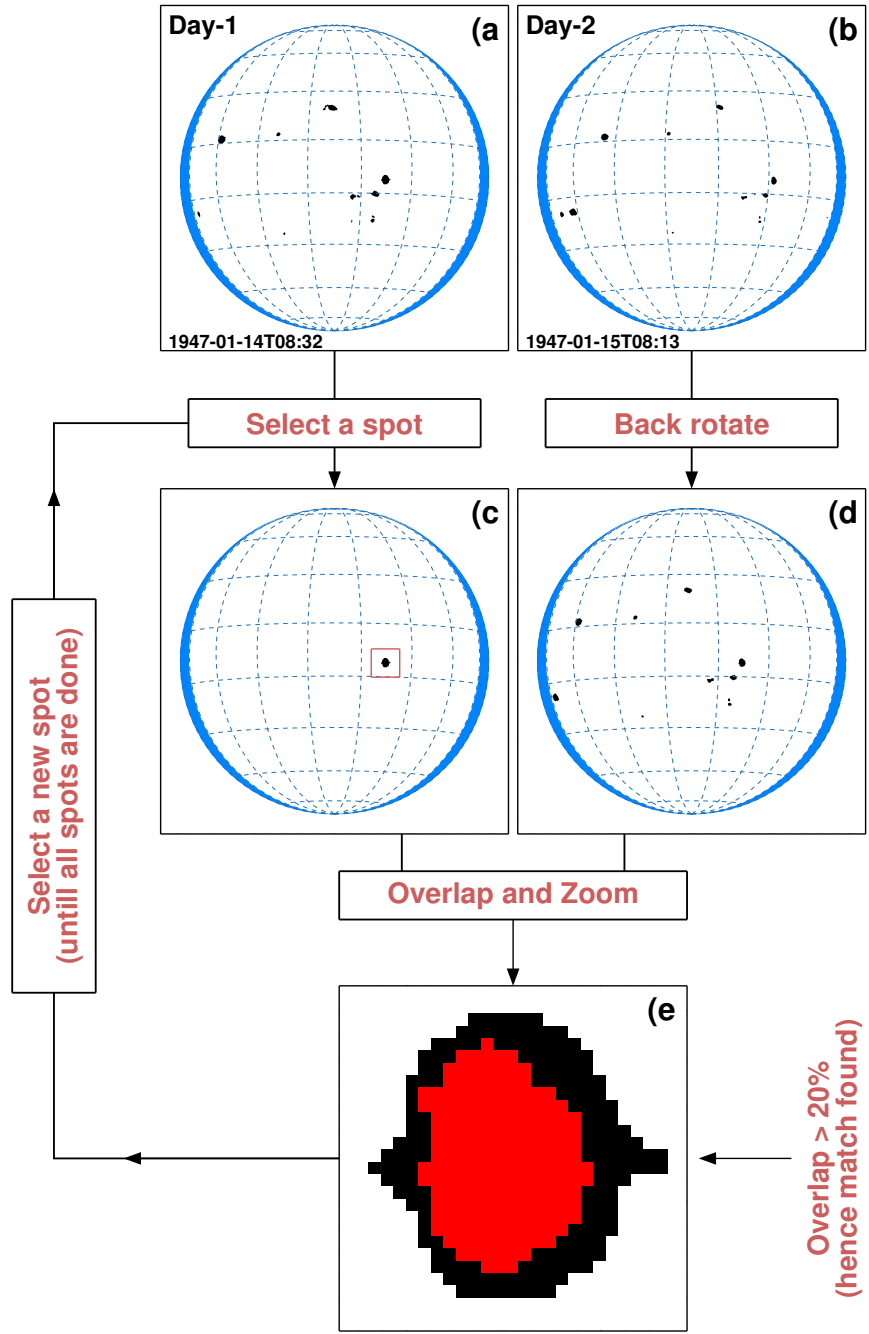}}
\caption{Representative examples of our sunspot tracking algorithm. (a), (b): Show two consecutive observations which are one day apart. Blue shaded regions near the solar limb highlight areas beyond $\vert 70\vert\degree$ longitudes.
(c) Shows a selected spot from Day-1 and (d) shows a back rotated image from Day-2.
(e) Depicts a zoomed in view of the overlapped region of Image1 (with the selected spot) on to the differentially back rotated image of Day-2.}
\label{fig1}
\end{figure}
%============================
To follow the same sunspot in two consecutive observations, we first choose two images, Image1 and Image2, which are (preferably) taken on consecutive days. Figures~\ref{fig1}a-b, show one such pair of observations. To incorporate the occasional `missing data' scenario, we also allow the time difference between Image1 and Image2 to be a maximum of up to three days. Additionally, to minimise the projection errors, we restrict our analysis to spots whose absolute heliographic longitude are $<\vert 70\vert\degree$.
In order to identify a selected spot of Image1 (e.g. \Fig{fig1}c) in a subsequent observation, we first differentially back rotate Image2 (using the IDL routine {\sf drot\_map.pro}\footnote{Detail of this function is available at \url{https://hesperia.gsfc.nasa.gov/ssw/gen/idl/maps/drot_map.pro}. This routine  uses the differential rotation parameter from \citet{howard_1990} to differentially rotate an input image.}) to the time of observation of Image1 (\Fig{fig1}d). In this context, it is also worth noting that the opposite sense of image rotation, i.e. forward-rotating Image1 to match the time of Image2 and repeat the same process also produces the same result.
Next, the selected spot from Image1 is overlapped onto the back rotated Image2 for the final step of the procedure (\Fig{fig1}d).
Basically, an overlap here is an indication of a potential match. However, group evolution makes this process somewhat complicated. To eliminate any false detection, we employ a two step verification method. Firstly, the overlap (by pixel count) has to be more than 20\%. This accounts for the uncertainties that are introduced during the back rotation of Image2. The example shown in \Fig{fig1}d has an overlap of 52\%. Next, there are situations when spots appear significantly different on Day 2 due to their rapid evolution (which are mostly mergers or bifurcations). To handle these cases, we invoke a cut-off on the fractional change in area as $\frac{|~a_2-a_1~|}{a_2+a_1} < 0.7$, where $a_1$ and $a_2$ are the areas of a spot in Image1 and Image2. This limit (of 0.7) has been decided after manually inspecting several randomly chosen image sequences. An Image1 spot which passes both these tests, is then flagged as the `same spot'. This whole procedure is repeated for all the spots in Image1 and for all images in our archive. 
In Appendix~\ref{appen}, we present a summary of our spot tracking algorithm in the form of a flow chart.

If $\phi _1$ and $\phi _2$ represent the heliographic longitudes of a spot at times $t_1$ and $t_2$, then the synodic rotation rate
($\Omega _{\rm synodic}$) is calculated as
\begin{equation}
     \Omega _{\rm synodic}~=~\frac{\phi _2 - \phi _1}{t_2 - t_1},
\end{equation}
to convert the synodic values into sidereal rotation rate, we use the following relation \citep{rosa_1995,wittmann_1996, soko_2014}:
\begin{equation}
    \Omega _{\rm sidereal}~=~\Omega _{\rm synodic}+ 0.9856\left(\frac{\cos{\psi}}{r^2}\right)\cos{i},
\end{equation}
where $i$ is the inclination of the solar equator to the ecliptic, $\psi$ is the angle between the pole of the ecliptic and the solar rotation axis orthographically projected on the solar disk and $r$ is the Sun-Earth distance in astronomical units \citep{lamb_2017}. We apply the above mentioned procedure on all the sunspots present in our data and calculated $\Omega _{\rm sidereal}$ (hereafter $\Omega$) values.

%========================Results===================================
\section{Results}
\label{s-results}

\subsection{The Average Rotation Profile}

\begin{figure}[!htbp]
\centerline{\includegraphics[width=0.8\textwidth,clip=]{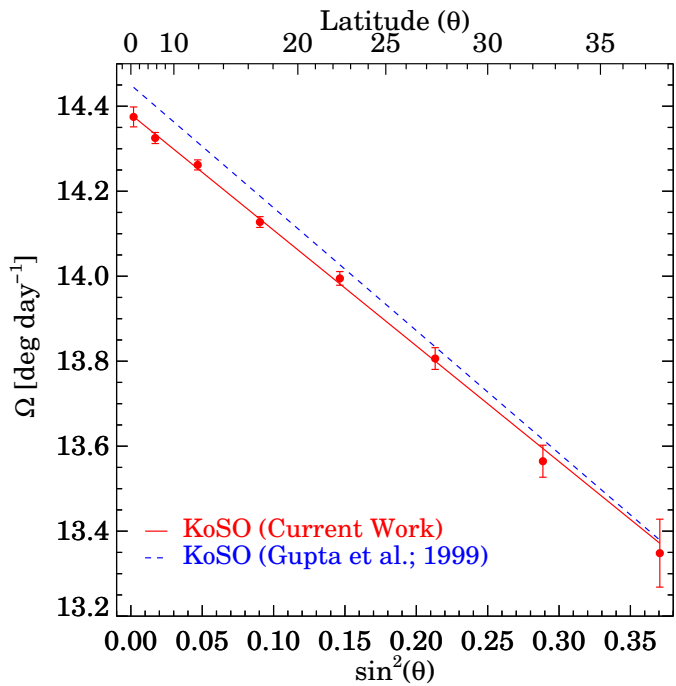}}
\caption{Solar rotation profile (red solid line) measured using the KoSO sunspot data for the period 1923--2011. For a comparison, we overplot (blue dashed line) the results from \citet{gupta1999}}
\label{fig2}
\end{figure}

%=========================================================================================================
\begin{table}
\begin{tabular}{ ccccc }
\hline
     References  & Period  & $A \pm \Delta A$ & $B \pm \Delta B$ \\
      & & ($\deg$$/$day) & ($\deg$$/$day) & \\
    \hline

    \citet{HG1984}         & 1921--1982 & $14.522  \pm 0.004$   & $-2.84 \pm 0.04$    \\
    \citet{howard1999a}
    \tabnote{Old low resolution KoSO data }
                            & 1907--1987 & $14.547\pm 0.005$     & $-2.96\pm 0.05$     \\
    \citet{howard1999a}
    \tabnote{Mt. Wilson data }
                           & 1917--1985 & $14.459\pm 0.005$     & $-2.99\pm 0.06$     \\
    \citet{gupta1999}
    \tabnote{Old low resolution KoSO data}
                          & 1906--1987  & $14.456 \pm 0.002$    & $-2.89 \pm 0.02$    \\
    \citet{ruzdjak_2017}
    \tabnote{Greenwich Photoheliographic Results and Debrecen Photoheliographic Data }
                          & 1874--2016  & $14.483\pm 0.005$     & $-2.67\pm 0.05$     \\

    Current Work
                           & 1923--2011 & $14.381\pm 0.004$        & $-2.72\pm0.04$  \\
\hline
\end{tabular}
\caption{Differential rotation parameters measured from different observations since \citeyear{carrington1863}.}
\label{table1}
\end{table}
%====================================================================================================================

To study the latitudinal dependency of solar rotation rate, we first organise the spots according to their latitudes ($\theta$) in $5\degree$ bins and calculate the mean $\Omega$ in each of those bins. In \Fig{fig2}, these mean $\Omega$ values (indicated by filled red circles) are plotted as a function of $\sin^2\theta$. The error bars shown in the plot are the standard errors calculated for each bin. As seen from the figure, $\Omega$ seems to have a linear relation with $\sin^2\theta$ \citep{howard1999a}. To measure the equatorial rotation rate ($A$) and the latitudinal gradient of rotation ($B$), we fit Equation~\ref{eq2} onto our data by using the Levenberg-Marquardt least squares (LMLS) method\footnote{We have used the {\sf mpfit\_fun.pro} function available in IDL for this purpose.} \citep{LMLS_2009} and, the obtained best fit is shown by the solid red line in \Fig{fig2}. The $A$ and $B$ values, returned from the fit, are $14.381\pm0.004$ and $-2.72\pm0.04$ respectively, and they compare well with the existing literature as shown in Table~\ref{table1}.
We directly compare our results with the values from \citet{gupta1999} (shown by the blue dashed line in \Fig{fig2}) who had used an older low-resolution version of the Kodaikanal data, of limited period as well. Interestingly, our $A$ and $B$ values are slightly different from their measurements (see Table~\ref{table1}) and we attribute such differences to the following two reasons. Firstly, \citet{gupta1999} used only the umbra to measure the rotation rate as opposed to the the whole-spot area (i.e. umbra and penumbra) as used in our study. Secondly, the epochs covered by these two datasets are different. As we will find in later sections not only the spatial extend of the tracer (i.e. sunspot) affects the derived $B$ value, but both parameters, $A$ and $B$, vary significantly with solar cycles. Thus, a combination of these two effects produces the observed differences.

%========================================================================================
\subsection{Variations in Rotational Profile}

\subsubsection{Sunspot Sizes}
\label{sizes}
\begin{figure}[!htbp]
\centerline{\includegraphics[width=0.7\textwidth,clip=]{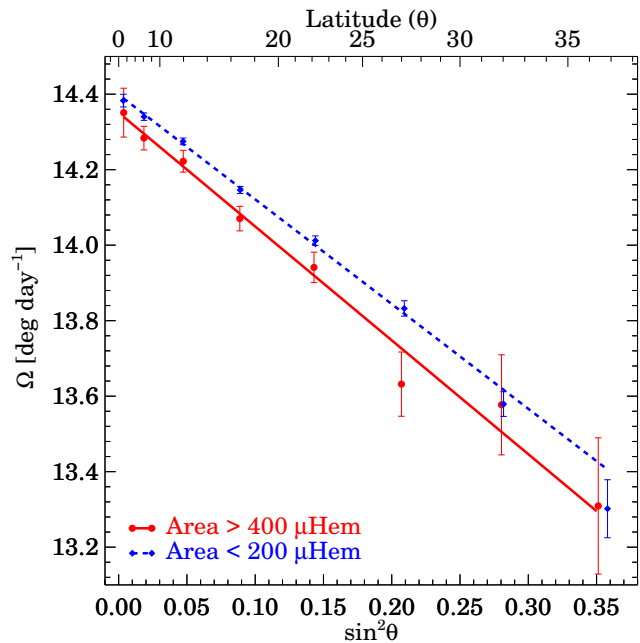}}
\caption{Rotation profiles of sunspots with area $<$200~$\mu$Hem (blue dashed line) and with area $>$400~$\mu$Hem (red solid line).}
\label{fig3}
\end{figure}

Sunspots are observed in a variety of shapes and sizes and from past observations, we know that bigger spots typically host stronger magnetic field \citep{jaramillo2015}. These strong fields, which are believed to be anchored deep in the convection zone, may have a possible effect on the spot rotation at the surface \citep{ward1966, GH1984, gupta1999}. Hence, we investigate the effect of stronger fields on the inferred solar rotation by dividing all the spots according to their areas. They are grouped into two area categories:  (i) small spots with area  $<200~\mu$Hem and, (ii) large spots with area  $>400~\mu$Hem. \Fig{fig3} shows the obtained rotation profiles for these two spot area classes.

Our results show that bigger spots rotate slower than smaller ones (see the $A$ values listed in Table~\ref{table2}). This is consistent with previous reports by \citet{ward1966, GH1984, gupta1999}; and \citet[][using magnetic field maps]{Kutsenko2020}. At the same time, we do not observe any significant change (considering the error limits) in $B$ values between the two area categories. Physically, the slower rotation rates for bigger spots (hence, stronger magnetic fields; \citealp{livingston2006}) may hint towards deeper anchoring depths of the parent flux tubes \citep{balthasar_1986}. Additionally, it has been pointed out that spots with larger areas experience a greater drag which further affects their rotation rates \citep{ward1966, GH1984}.

\begin{table}[h]
\begin{tabular}{ccccc}
    % \multicolumn{1}{c}{Data}&
    % \multicolumn{2}{c}{Kodaikanal}\\
     \hline
     Area Group  &$A\pm \Delta A$ &$B \pm \Delta B$ \\%&$A\pm \Delta A$ &$B\pm \Delta B$\\
    \hline
    Small (Area $<200~\mu$Hem) &$14.399\pm 0.004$        &$-2.77 \pm 0.04$\\      %   
    Large (Area $>400~\mu$Hem) &$14.351\pm 0.011$        &$-3.01 \pm 0.12$   \\   %  & \
    \hline
\end{tabular}
\caption{Differential rotation parameters measured from different sizes of spots for Kodaikanal  Solar Observatory data (current work).}
\label{table2}
\end{table}

\subsubsection{Solar Activity Phase and Cycle Strength}
Properties of sunspots such as their areas, numbers, etc. vary with solar cycle phases \citep{Hathaway2015}. Hence, it is reasonable to look for signatures of any such variation in the $\Omega$ profile.

\begin{figure}[!htbp]
{\centerline{\includegraphics[width=\textwidth,clip=]{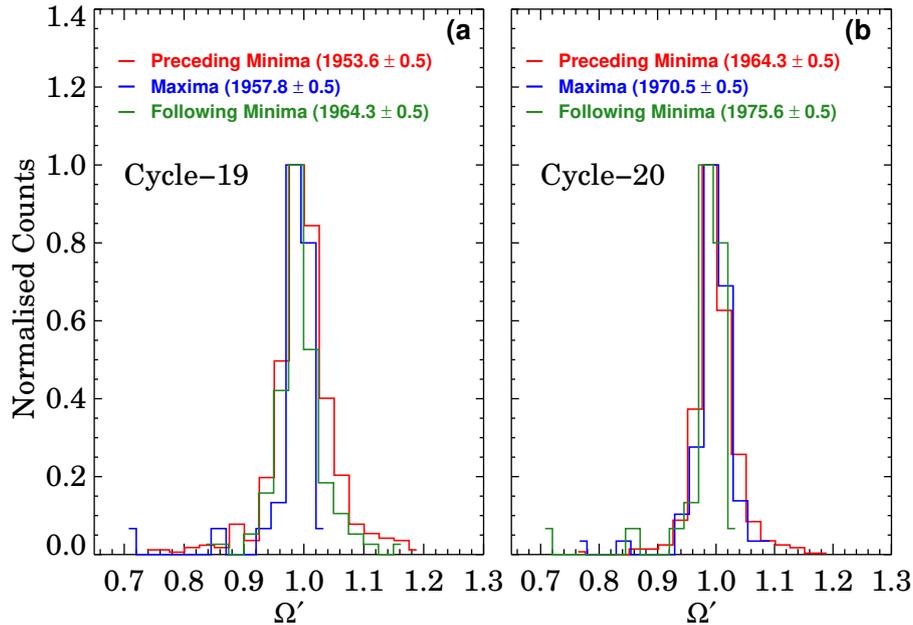}}}
\caption{(a): Distributions of $\Omega^\prime$, for Cycle 19, as calculated over the three activity phases: preceding minima (red), maxima (blue), and following minima (green). See text for more details. (b): Same as in (a), but for cycle 20.}
\label{fig4}
\end{figure}

For every solar cycle, we first isolate the data into three activity phases: preceding minimum, cycle maximum, and following minimum. These phases (each with one year duration) are identified after performing a 13 month running average on the KoSO area data. Next, we choose a 30$\degree$ latitude band centred around the solar equator. Choice of such a bandwidth ensures that we have enough statistics to work with during each of these phases. Since the average size of spots varies with the phase of the solar cycle and as shown in Section~\ref{sizes}, spots with different sizes rotate with different rates, it is important to remove this effect from the measured $\Omega$ before we look for its temporal variation. We achieve this by dividing the calculated $\Omega$ values of every spot by the $\Omega$ derived using the rotation parameters $A$ and $B$ from Table~\ref{table2}. This newly normalised quantity is denoted by $\Omega^\prime$ and the distributions of $\Omega^\prime$, for all three activity phases and for all eight solar cycles, are then examined thoroughly. \Fig{fig4} shows the results for two representative cases, i.e for Cycle 19 and Cycle 20. As noted from the plot, distribution of $\Omega^\prime$ shows no change, either in shape or location, with activity phases. Table~\ref{table_md} lists the statistical parameters (such as mean, skewness, etc.) of each of these distributions which further confirm our previous observation. Overall, our results are in accordance with the findings of \citet{GH1984}, \citet{ruzdjak_2017} and \citet{Javaraiah2020}, who either found none or statistically insignificant correlation of sunspot rotation with the activity phase of the solar cycle.

\begin{table}[!h]
\begin{tabular}{crrrrrr}
\hline
    \multicolumn{1}{c}{}&
    \multicolumn{3}{c}{Maximum}&
    \multicolumn{3}{c}{Following Minimum}\\
\hline
Cycle &Mean &Median &Skewness &Mean &Median &Skewness\\
\hline

    16 &$     1.000$ &$     0.998$ &$     0.190$ &$     1.000$ &$     0.998$ &$     0.010$ \\
    17 &$     1.002$ &$     0.999$ &$     0.279$ &$     0.993$ &$     0.991$ &$     0.521$ \\
    18 &$     0.996$ &$     0.995$ &$    -0.030$ &$     0.991$ &$     0.992$ &$    -0.042$ \\
    19 &$     1.000$ &$     0.998$ &$     0.035$ &$     0.993$ &$     0.993$ &$    -0.027$ \\
    20 &$     0.998$ &$     0.996$ &$     0.188$ &$     0.997$ &$     0.997$ &$    -0.272$ \\
    21 &$     1.003$ &$     1.000$ &$     0.275$ &$     0.995$ &$     0.992$ &$    -0.079$ \\
    22 &$     0.997$ &$     0.996$ &$    -0.053$ &$     0.998$ &$     0.997$ &$    -0.054$ \\
    23 &$     1.000$ &$     0.997$ &$     0.203$ &$     1.010$ &$     1.008$ &$     0.577$ \\

\hline
\end{tabular}
\caption{Statistical parameters of $\Omega^\prime$ distributions for each cycle.}
\label{table_md}
\end{table}

\begin{figure}[!htbp]
\centerline{\includegraphics[width=0.8\textwidth,clip=]{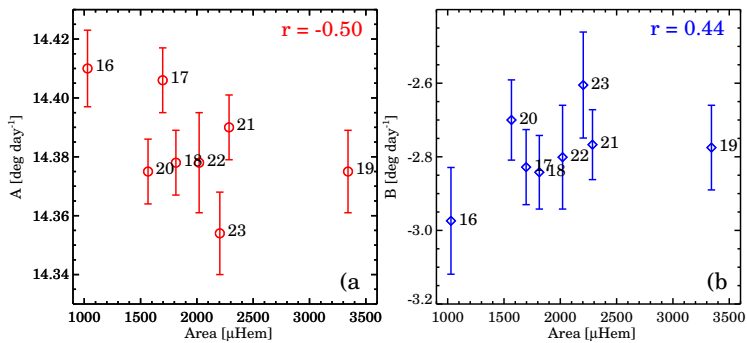}}
\caption{(a): Scatter plots between $A$ and cycle strengths. (b): Scatter plots between $B$ and cycle strengths. Obtained Pearson correlations ($r$) are printed in the respective panels.
}
\label{fig5}
\end{figure}

Next, we explore how a rotational profile depends upon the strength of a solar cycle. We do this by analysing the rotation parameters $A$ and $B$ of each cycle. The peak area value within a cycle (of yearly averaged data) is assigned as the strength of that cycle. In Figure~\ref{fig5}, we show the scatter plot of $A$ vs. cycle strength (Figure~\ref{fig5}a), as well as the plot between $B$ vs. cycle strength (Figure~\ref{fig5}b). Immediately we notice that $A$, which represents the equatorial rotation rate, decreases with increasing cycle strengths (Pearson correlation, $r=-0.50$). Interestingly, $B$, the latitudinal gradient of rotation, shows a positive correlation ($r=0.44$). This implies that, during a strong cycle, the Sun not only rotates slowly at the equator but the latitudinal gradient of rotation also gets reduced \citep{GH1984,Obridko2001,java2005, Obridko2016, Javaraiah2020}. It is important to highlight here that in both panels, we notice that the values for Cycle 19 seem to lie significantly far away from the trends. At this moment, we do not know any physical process that can explain this observed anomaly.

\subsubsection{Rotation Rates of Northern and Southern Hemispheres}

Almost all sunspot indices display profound differences in their properties, in the northern and southern hemisphere \citep{Hathaway2015}, including sunspot rotation rates \citep{GH1984}. However, in the case of rotation rates, the difference, as reported by many authors in the past, differs significantly from each other. For example, \citet{GH1984} used sunspots as tracers and noted that the variation in solar rotation is more profound in the southern hemisphere relative to the northern one. In a recent work, \citet{Xie_2018} reported that the northern hemisphere rotates considerably faster than the southern one during Cycles 21\,--\,23. Similar findings, for the same period, have been also reported earlier \citep{zhang2011, zhang2013, li2013a}. Thus, comparing the rotation rates in both hemispheres from our KoSO sunspot data is appropriate.

\begin{figure}[!htbp]
{\centerline{\includegraphics[width=0.8\textwidth,clip=]{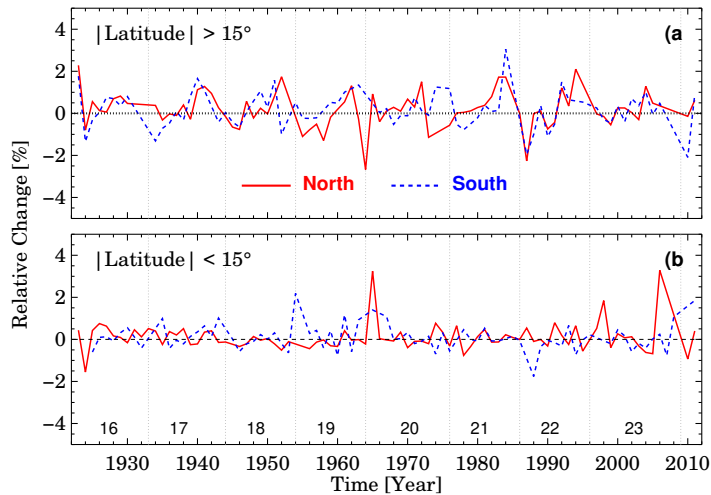}}}
\caption{Relative (\%) changes in $\Omega$, separately for northern and southern hemispheres, in latitude band~I (b) and band~II (a). See text for more details. }
\label{fig6_}
\end{figure}

To achieve this, the data is first separated for the two hemispheres and then each hemispheric data is divided into two separate latitude bands: band~I, 0\,--\,15$\degree$ and band~II, $\geq$15$\degree$. In each case, we then calculate the relative change by first calculating the mean $\Omega$ in that band over the year ($\bar{\Omega}_{\rm Year}$) and then calculate the relative change as $\frac{\bar{\Omega}_{\rm Year}-\bar{\Omega}_{\rm All}}{\bar{\Omega}_{\rm All}}\times 100\%$, where $\bar{\Omega}_{\rm All}$ is the mean $\Omega$ calculated over the entire duration of the data (1923\,--\,2011) within our chosen band.
Figure~\ref{fig6_}a-b show these relative changes for band~II and band~I, respectively. In the case of band~I, the relative change is $<$0.5\% and we do not observe any signature of solar cycle like variation. On the other hand, in the case of band~II, there are clear signatures of cyclic changes (e.g. for Cycles 16, 17, 18, 21) along with a slightly larger variation ($\approx$ 1.5\%) as compared to band~I.  

\section{Cross-Validation of Our Method Using Space Based Data}

As discussed in the introduction, rotational profiles that have been derived using sunspots, depend significantly (i) on the method used, and (ii) on the quality of the data. To examine the aforementioned effects onto our measurements of rotation rates from KoSO data, we use the space based sunspot data available from MDI onboard SOHO. These data cover a period of 15 years (1996\,--\,2011). 

\begin{figure}[!htbp]
\centerline{\includegraphics[width=\textwidth,clip=]{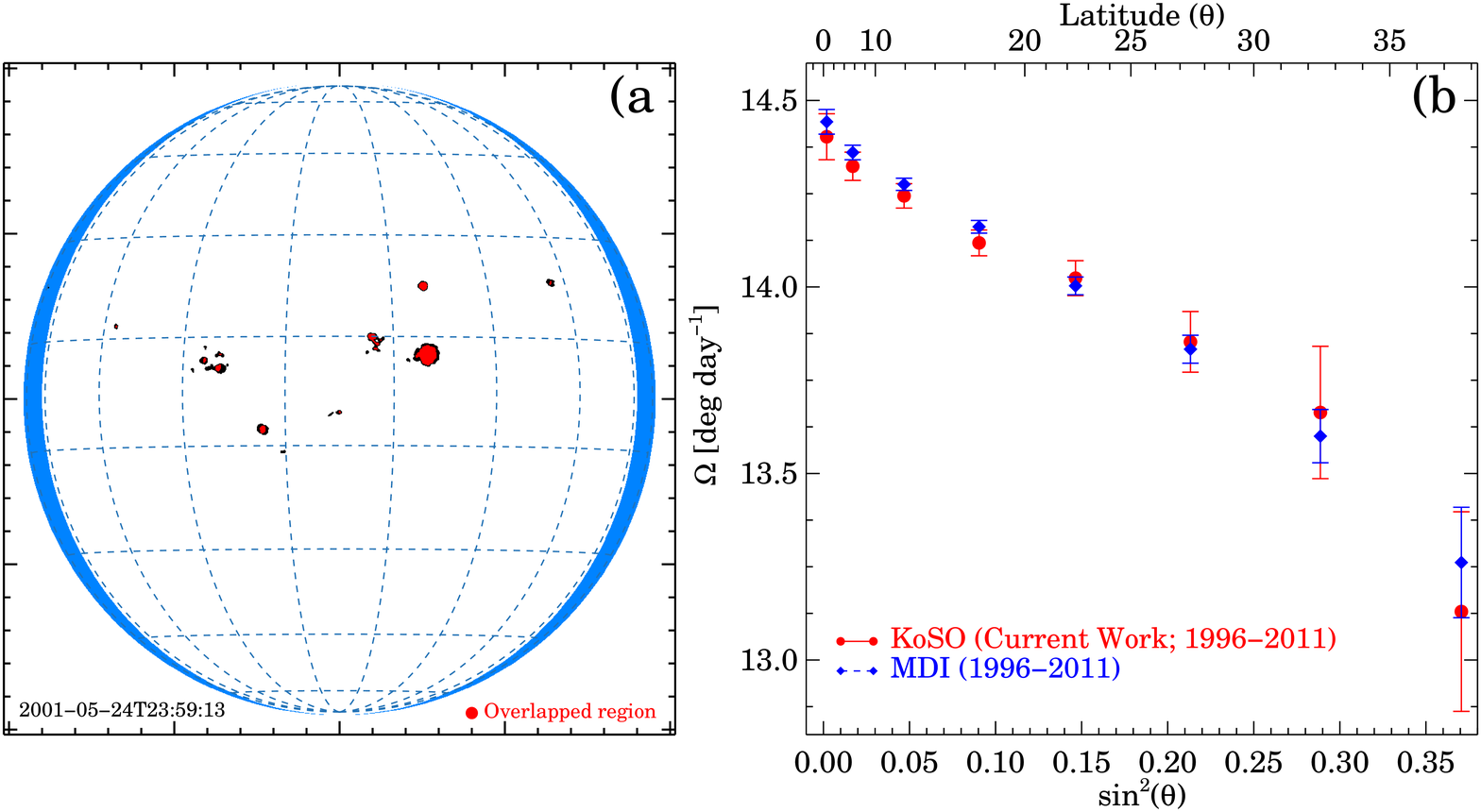}}
\caption{(a): Representative MDI image after implementing our sunspot tracking algorithm. (b): Solar rotation profile measured using MDI data (blue circles) and the rotation profile derived from KoSO data for the overlapping period (red circles).
}
\label{fig7}
\end{figure}

Sunspot masks are generated using the same methods as used in KoSO data and likewise, the method implemented to track the spots (Figure~\ref{fig7}a) and derive the rotation profile, is also the same as described in Section~\ref{s-method}. Results from MDI are shown in Figure~\ref{fig7}b using blue circles. For easy comparisons, results from our KoSO data (corresponding to the same epoch) are also overplotted (via red circles) in both panels. As seen from the plots, the KoSO profile matches well with the values from MDI. At the same time, we also note that the error bars are substantially smaller in MDI which is mostly due to the better data availability in MDI as compared to KoSO within this period. Overall, these results not only show the effectiveness of our method on ground and space based data but also highlights the quality of the KoSO catalogue.

%=======================================================================================

%========================================================================================
\section{Conclusion}
\label{s-con}

A better knowledge of solar differential rotation is key towards a deeper understanding of solar dynamo theory. In this work, we have derived the solar rotation profile using the white-light digitised data from Kodaikanal Solar Observatory (KoSO). These data cover a period of $\approx$90 years (1923\,--\,2011) i.e. Cycle~16 to Cycle~24 (ascending phase only). We have implemented a correlation-based automated sunspot tracking algorithm which correctly identifies the same sunspot in two consecutive observations. This positional information is then used to calculate the angular rotation rate for every individual spot. Furthermore, the derived rotation parameters $A$ and $B$ ($14.381\pm0.004,~-2.72\pm0.04$; Figure~\ref{fig2}), which represent the equatorial rotation rate and the latitudinal gradient of rotation, respectively, are in good agreement with existing reports from other sunspot catalogues \citep{java2005, Ribes1993}. In fact, our results also compared well with results from the older, low-resolution, and manually processed KoSO sunspot data \citep{howard1999a, gupta1999}.

During our analysis, we found that bigger spots rotate slower than smaller ones (Figure~\ref{fig3}), which could well be a possible signature of deeper anchoring depths of bigger spots \citep{balthasar_1986}. We have also studied the cyclic variations in the rotation profile along with its modulation with cycle strengths. Results suggest that during a strong cycle, the Sun not only rotates slower at the equator but the latitudinal gradient of rotation also gets reduced (Figure~\ref{fig5}). Another interesting finding from this study is the relative changes in the rotation rate in the northern and southern hemispheres, where high latitude ($>$15$\degree$) spots seem to show pronounced variations ($\approx$1.5\% as compared to low latitude ($<$15$\degree$) ones (\Fig{fig6_}). Lastly, we check for the robustness of our tracking algorithm by implementing it onto the MDI data. Similar results from KoSO and MDI confirm the same.

In this study we have assumed that sunspots are basically embedded structures within the photosphere and their motions, such as the rotation about their own axis \citep{Brown2003,svanda2009} is not accounted for in our study. The KoSO image database (with one day cadence and relatively low spatial resolution) is not really capable of taking care of these effects and we need modern day high resolution (spatial and temporal) data for such a study.
Furthermore, KoSO has historical data for the last century in two other wavelengths, i.e. H$\alpha$ and CaII K. These datasets can further be used in the future to investigate the change in solar rotation above the photospheric height and in turn, this will provide insight about the coupling between different solar atmospheric layers.

%====================================================================================
\begin{acks}
{Kodaikanal Solar Observatory is a facility of Indian Institute of Astrophysics, Bangalore, India. These data are now available for public use at http://kso.iiap.res.in through a service developed at IUCAA under the Data Driven Initiatives project funded by the National Knowledge Network. We would also like to thank Ravindra B. and Manjunath Hegde for their tireless support during the digitisation, calibration, and sunspot detection processes. We will also thank Ritesh Patel and Satabdwa Majumdar for fruitful discussions during the work}
\end{acks}

\paragraph*{\footnotesize Disclosure of Potential Conflicts of Interest}
The authors declare that they have no conflicts of interest.
%=====================================================================
\newpage
\appendix
\section{Flow Chart}
\label{appen}
\begin{figure}[!htbp]
\centerline{\includegraphics[width=\textwidth,clip=]{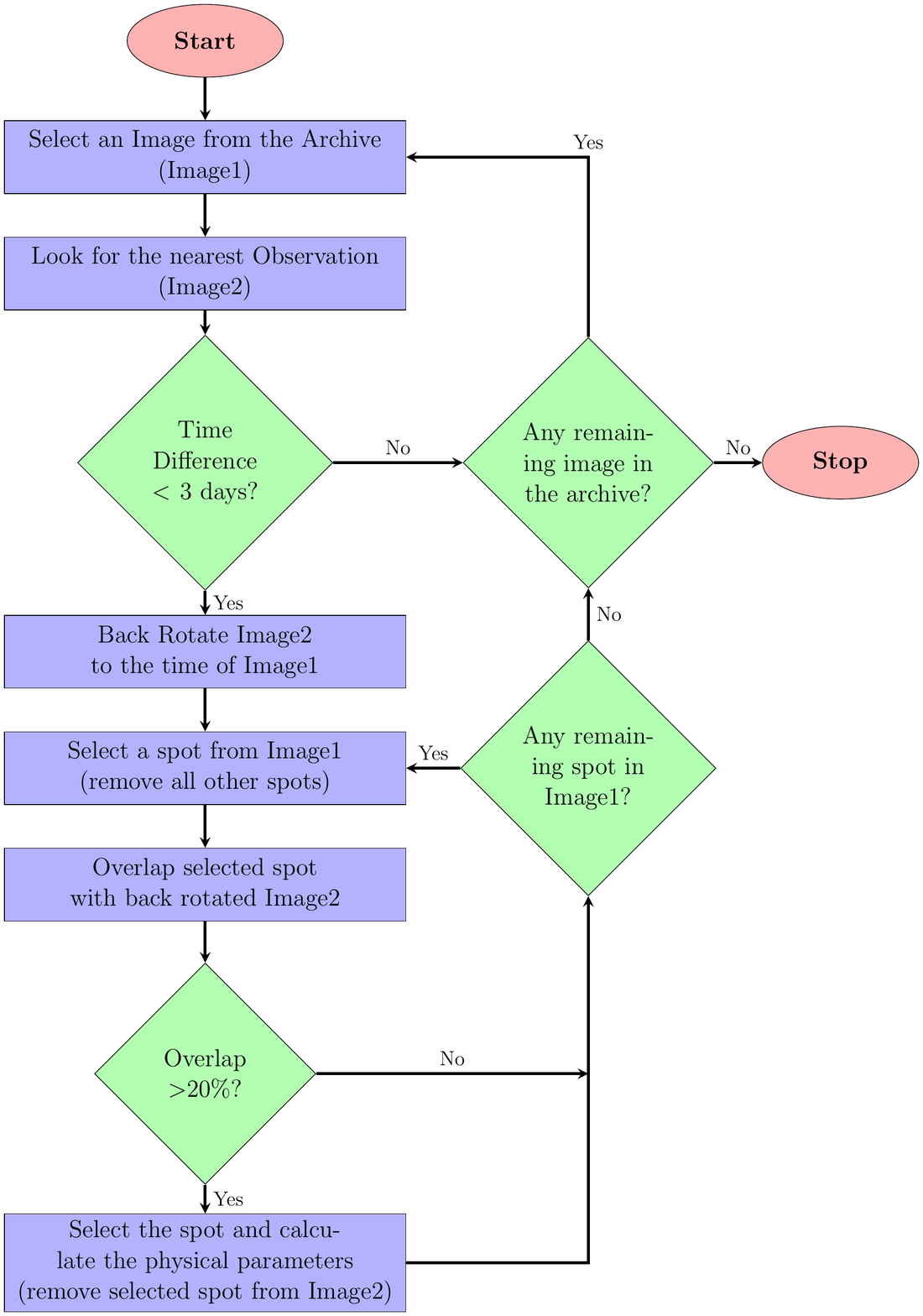}}
\caption{A flow chart describing the different steps of our spot tracking algorithm.}
\label{appn_figure}
\end{figure}
%==================================================================================

\newpage
\bibliographystyle{spr-mp-sola}
%\bibliography{diff_2019}  

% Checking: look if the file containing the ``\bibitem'' exits
%           so check if the .bbl file exist (bibTeX compilation)
 \IfFileExists{\jobname.bbl}{} {\typeout{}
 \typeout{****************************************************}
 \typeout{****************************************************}
 \typeout{** Please run "bibtex \jobname" to obtain} \typeout{**
 the bibliography and then re-run LaTeX} \typeout{** twice to fix
 the references !}
 \typeout{****************************************************}
 \typeout{****************************************************}
 \typeout{}}

%=====================================================================

%=====================================================================

\end{article}

\end{document}